# Why the Model of a Hydrogen-filled Sun is Obsolete

O. Manuel, Nuclear Chemistry, U. Missouri-Rolla

## I. Historical Background

Noting that the Earth's crust and the Sun's gaseous envelope may not represent the overall compositions of these bodies, Harkins[1] showed in 1917 that 99% of the material in ordinary meteorites consists of seven, even-numbered elements - iron (Fe), oxygen (O), nickel (Ni), silicon (Si), magnesium (Mg), sulfur (S) and calcium (Ca). He concluded that *"... in the evolution of elements much more material has gone into the even-numbered elements than into those which are odd..."* (p. 869). However, in the 1920s, Payne[2] and Russell[3] showed that the solar atmosphere is mostly hydrogen (H) and helium (He).

In 1938 Goldschmidt[4] suggested that rocky planets and meteorites had lost volatile elements such as hydrogen and helium. He proposed an abundance table based on the solar atmosphere.

Hoyle[5] maintained, however, that the results of Payne[2] and Russell[3] were for the atmospheres, not for the deep interiors of stars. He, Eddington, and other astronomers continued to believe until the end of World War II that *"... the Sun was made mostly of iron ..."* (p. 153). Research on hydrogen-fusion at Los Alamos during the war[6] likely aided their conversion.

Goldschmidt's proposed loss of light, volatile elements from meteorites and rocky planets[4] seems reasonable from the view of planetary evolution. From the view of nuclear physics the difference between an iron-rich and a hydrogen-rich Sun is, however, drastic.

Iron (Fe) is an even-numbered element. It consists mostly of the intermediate-mass isotope, $^{56}$Fe. This has an ordinary charge density, (atomic number)/(mass number) = $Z/A$ = 0.46. Its nucleons have the lowest[7] known potential energy, 1.082 million electron volts (MeV) less than the average nucleon energy in the most abundant carbon isotope, $^{12}$C.

Hydrogen is an odd-numbered element. It consists mostly of the lightest isotope, $^{1}$H, with the highest charge density ($Z/A$ = 1.00). Its nucleon has the highest[7] potential energy (M/A) among stable nuclides, 7.289 MeV more than the average nucleon energy in $^{12}$C.

The hydrogen-rich Sun with its inherent nuclear instability thus violates, on a grand scale, Harkins' prediction[1] (p. 859) that *"... the more stable atoms should be more abundantly formed ..."*. However, adherents to the hydrogen-filled Sun model (e.g., Suess & Urey[8]) continued to use nuclear stability to try to explain the abundance of elements composed of nucleons with intermediate potential energies (MeV).

To try to explain the existence of a hydrogen-rich Sun, Burbidge et al.[9] and Cameron[10] assumed that the products of nucleosynthesis were mixed back into the hydrogen-rich interstellar medium before the solar system formed. The 1960 discovery[11] of radiogenic xenon-129 from the decay of extinct iodine-129 in meteorites left little time for mixing[12], and this mixing time became impossibly short as high sensitivity mass spectrometers[13] later revealed the decay products of even shorter-lived nuclides and nucleogenetic isotopic anomalies in numerous elements that comprise meteorites.

My work on this project began in 1960 when, as a graduate student, I joined Professor Paul Kuroda's research group at the University of Arkansas. It has continued to date. My conclusion of an iron-rich Sun appeared in a 1998 review[14] and in the proceedings of an ACS symposium[15] organized by Seaborg and Manuel in August, 1999. The experimental evidence is summarized in Section II. The conclusion that a hydrogen-filled Sun is obsolete, related observations, and the status of work on remaining issues are given in Section III.

## II. Evidence for an Iron-Rich Sun

Samples collected by the *Apollo* missions to the moon in the late 1960s and early 1970s revealed that lighter mass ($m_L$) isotopes of helium (He), neon (Ne), argon (Ar), krypton (Kr), and xenon (Xe) are enriched in the solar wind (SW) relative to the heavier ($m_H$) ones by a common mass-fractionation factor[15,16] [See p. 281*], $f$, where

$$f = (m_H/m_L)^{4.56}$$

When this empirical power law, defined by enrichments of light isotopes in the solar wind, was applied to solar atmospheric abundance, the most abundant elements in the Sun were found[16] to be iron (Fe), nickel (Ni), oxygen (O), silicon (Si), sulfur (S), magnesium (Mg), and calcium (Ca) [p. 283]. These elements all have even atomic numbers, they are made in the interior of supernovae[9], and they are the same seven elements Harkins[1] found in 1917 to comprise 99% of ordinary meteorites.

Could this be a coincidence? Hardly! If all 83 elements in the Sun's atmosphere were equal in abundance, the probability for the chance selection of any set of seven elements would be 7! 76! /83! = 2 x $10^{-10}$. This differs little from zero. The actual probability for chance selection of these seven elements is orders-of-magnitude less, in fact less than 2 x $10^{-33}$ because the probability of selection would depend on the abundance of each of these trace elements in the solar atmosphere. Clearly the Sun's most abundant elements are iron (Fe), nickel (Ni), oxygen (O), silicon (Si), sulfur (S), magnesium (Mg), and calcium (Ca).

Light isotopes of helium (He), neon (Ne), magnesium (Mg), and argon (Ar) are systematically less enriched in solar flares as if these energetic events by-pass 3.4 stages of mass-fractionation[17] [p. 282]. Measurements with the *Wind* spacecraft recently confirmed[18] that heavy elements are methodically enriched in material ejected by impulsive solar flares. The prevalence of solar wind implanted $^6$Li and $^{10}$Be in lunar soils is too high to be representative of the composition of the entire Sun according to the authors[19,20].

Linked isotopic and elemental variations[21,22] in meteorites first hinted that elements from deep in a supernova formed the interiors of the Sun and the terrestrial planets [pp. 593, 601]. Primordial helium from the outer layers of the supernova accompanies "strange" xenon[23], with excess $^{136}$Xe and $^{124}$Xe, but not "normal" xenon from inside[21,22,24,25] the supernova [p. 603]. Support for a supernova origin of the solar system [p. 593] came from findings of a) excess r- and p-products in other heavy elements[26] that accompany the "strange" xenon [p. 361]; b) complementary isotopic components enriched in s-products[27] [pp. 380, 619]; c) age dating[28] based on extinct [pp. 616-617] and longer-lived [pp. 490-491] nuclides; d) terrestrial-type xenon[29] in iron sulfide (FeS) of diverse meteorites, in planets like Earth and Mars that are rich in iron (Fe) and sulfur (S), and in the solar wind, where light isotopes are enriched by 3.5% per mass unit[16] [p. 623]; and e) despite poor quality data from the *Galileo* mission to Jupiter, the finding[30] of "strange" xenon in Jupiter's helium-rich atmosphere [pp. 519-527] and isotopes of hydrogen (H) and helium (He) in Jupiter that could not be converted into those seen in the solar wind by deuterium-burning [pp. 529-543].

### III. Conclusions, Related Observations, and Remaining Problems

The flux of neutrinos from the Sun is too low if one attributes solar luminosity entirely to the fusion of hydrogen. This has been a long-standing enigma[31] for adherents to the model of a hydrogen-rich Sun. The observations reported here confirm that that model is now obsolete. On the other hand, an iron-rich Sun that formed from supernova debris offers a direct explanation for these observations and for:
- heterogeneous accretion of terrestrial planets[32-34],
- primordial helium (He) and radiogenic xenon-129 ($^{129}$Xe) inside the Earth today[35,36],
- non-magmatic iron meteorites [pp. 385-406],
- isotopic anomalies and decay products of short-lived nuclides in iron [pp. 385-406] as well as in primitive [pp. 361-384] meteorites[37],
- the iron gradient in planets and in the planetary system [pp. 608-611], and
- experimental affirmation[38] of all three tests originally proposed to test the hypothesis that mass fractionation enriches lighter particles at the Sun's surface[16].

Thus, mass fractionation made the solar atmosphere that is 91% hydrogen (H, the lightest element), 8.9% helium (He, the next lightest element), and only about 0.1% of the 81 heavier elements that comprise the bulk of the Earth, the planets close to the Sun, and the meteorites (Fe, O, Si, Ni, Mg, S, Ca, etc.).

Two of three major objections to the iron-rich Sun have been resolved and a third is in progress. The discovery[39,40] of pulsar planets confirmed that planetary systems can form directly from supernova debris[41]. The Hubble telescope verified the existence of axial ejections from supernovae [pp. 241-249]. Extrapolations of trends from the cradle of the nuclides [book cover[15]] indicate an inherent instability in assemblages of neutrons that may explain solar luminosity and the solar neutrino flux[42,43].

After correcting for fractionation, solar abundance generally correlates with nuclear stability[43], as Harkins[1] had predicted in 1917, except for a large excess of the lightest hydrogen isotope, $^1$H. On Thursday, Cynthia Bolon will show that this anomalous $^1$H and the outflow of $^1$H+ ions in the solar wind are apparently by-products of solar luminosity[42-46].

Many of the experimental observations which form the basis for this paper can be viewed on the web at ***http://www.umr.edu/~om/***

———

*All page numbers in brackets [ ] are in ref. 15, Proceedings of the 1999 symposium organized by Glenn T. Seaborg and Oliver K. Manuel. This book is available from the publisher at the AAS meeting or on the web at ***http://www.wkap.nl/book.htm/0-306-46562-0***

This paper is in memory of my teacher, Professor Paul Kazuo Kuroda, who discovered the presence of extinct $^{244}$Pu and reported the $^{244}$Pu and $^{26}$Al ages shown on the last page for the formation and early history of the solar system. The last names of Kuroda's former students and successive generations --- students of Kuroda's former students, etc. --- are underlined in the reference list.

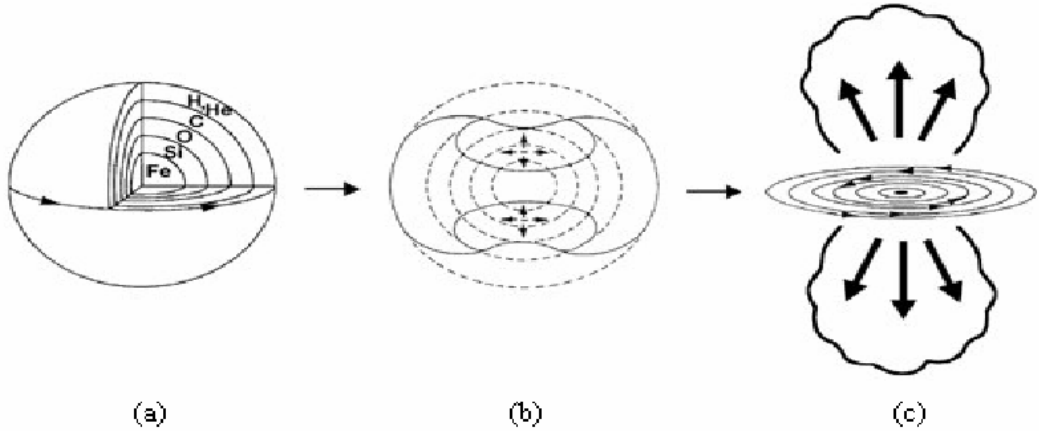

Fig. 1. The solar system formed directly from heterogeneous debris of a spinning supernova [26] that exploded axially to produce a bipolar nebula. (a) A massive spinning star becomes chemically layered near the end of its life; (b) Asymmetric collapse occurs to consume angular momentum and the in-fall of low-Z elements causes an axially directed explosion; (c) The sun forms on the SN core; cores of inner planets form in the iron – rich regions around the SN core; Jovian planets form from the outer SN layers.

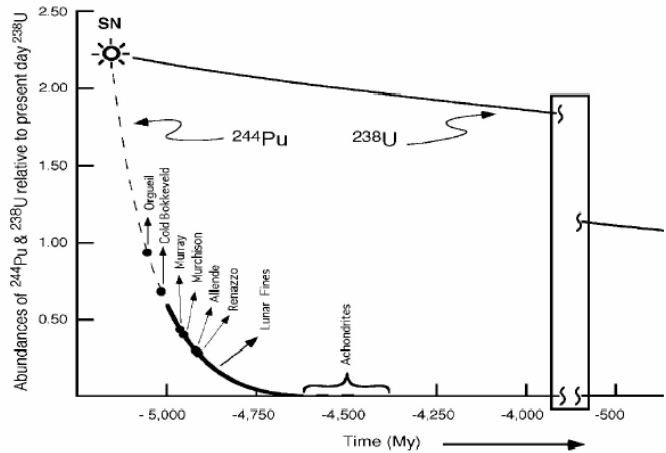

Fig 2. Carbonaceous chondrites start to retain gaseous fission products of Pu-244 about 100 My after the SN event [28].

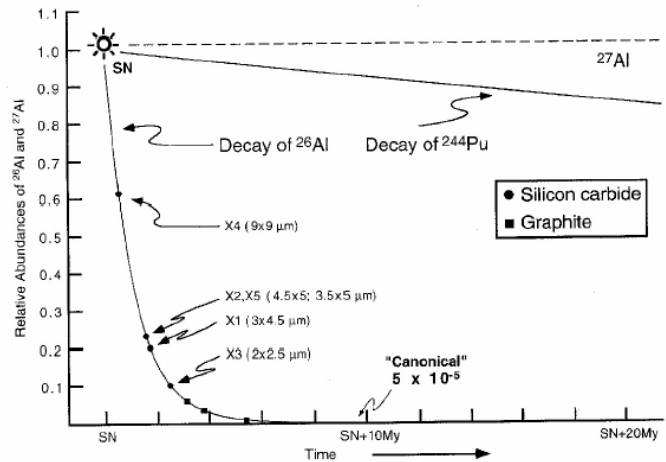

Fig 3. Radioactive Al-26 and large isotopic anomalies were trapped in graphite and in X grains of silicon carbide from the Murchison meteorite with a few My of the SN event [28].